\documentclass{article}
\title{Speed of light in the extended gravity theories}
\author{A. Izadi and A. Shojai
\\ Department of Physics, University of Tehran,
\\ North Karegar Ave., Tehran, Iran.}
\date{}
\def\az{\mathcal{R}}
\def\sz{\mathcal{G}}
\begin{document}
\maketitle
\begin{abstract}
We shall investigate the possibility of  formulation of  varying speed of light (VSL) in the framework of Palatini non-linear Ricci scalar and Ricci squared  theories.  Different speeds of light including the causal structure constant, electromagnetic, and gravitational wave speeds are discussed. We shall see that two local frames are distinguishable and discuss about the velocity of light in these two frames. We shall investigate which one of these local frames is inertial. 
\end{abstract}
\section{Introduction}
There are many theoretical and experimental suggestions that the values of fundamental \textit{physical constants} are not actually constant and may vary in space and time\cite{yek}. One of these \textit{constants} which has attracted considerable attention recently\cite{magmofasal} is the speed of light. The first modern VSL theory was Moffat's ground breaking paper \cite{moffat}, where spontaneous breaking of Lorentz symmetry leads to VSL theories. Then these theories followed by Albrecht, Magueijo and Barrow\cite{1khodam,2khodam} as an alternative to the inflation mechanism to solve some puzzles of Big-Bang cosmological models\cite{1khodam,3khodam}. 
In such a VSL cosmology the Lorentz invariance is explicitly 
broken and there is a preferred reference frame in which the physical laws should be formulated. They called this frame the cosmological frame and assumed that in this frame interactions have to be introduced via  a minimal coupling principle according to which the curvature tensor is computed ignoring the variation of the  speed of light. In this way the field equations are  Einstein's equations with the varying speed of light appearing in the right hand side.

Generally  in VSL models either a pre-set function for the speed of light is considered\cite{2khodam,4khodam} 
or there is a term in the Lagrangian determining the dynamics of the speed of light\cite{5khodam}.
However Magueijo in \cite{5khodam} has proposed a generalized varying speed of light theory preserving the general covariance and local Lorentz invariance. This is achieved by introducing a time-like coordinate $x^0$ which is not necessarily equal to $ct$, and the physical time $t$ can only be defined when $dx^0/c$ is integrable.

It has to be emphasized here that there are different constants which can be interpreted as the velocity of light. As Ellis and Uzan\cite{ellis} have shown one have to distinguish between $c_{EM}$ (the electromagnetic wave velocity), $c_{ST}$ (the space--time causal structure constant), $c_{GW}$ (the gravitational wave velocity) and $c_E$ (the space--time--matter coupling constant appearing at the right hand side of Einstein's equations). 
They have shown that assuming the standard Lagrangian of the electromagnetism and general relativity, one has $c_{EM}=c_{ST}$ and also that $c_{GW}=c_{ST}$. They have also shown that having the correct Newtonian limit leads to $c_E=c_{ST}$.

As a result, in dealing with any modified gravity and/or electromagnetism theory one should  re-examine the meaning and the relation between these concepts of the speed of light. Also to make the concept of varying speed of light meaningful, one should consider dimensionless quantities like $c_E/c_0$, $c_{GW}/c_0$, $c_{EM}/c_0$, and $c_{ST}/c_0$, where $c_0$ is a constant of dimension velocity and we can choose it equal to $3\times 10^8\textit{meters}/\textit{seconds}$ in the MKS units. In this way for the standard theory of electromagnetism and gravity we have: $c_E/c_0=c_{GW}/c_0=c_{EM}/c_0=c_{ST}/c_0=1$.

On the other hand, there are many different motivations to replace the standard general relativity by a more complete one which has richer space--time concepts. The existence of solutions with curvature singularities or closed-time-like loops in the standard general relativity are some of these reasons. Furthermore the flat rotation curves of galaxies and the current accelerated expansion of the universe introduce dark matter and dark energy with an amount about $96\%$ of the contents of the universe may be interpreted as signals for need for an extended theory. 

There are different approaches to modify Einstein's gravity theory. These can be put into different 
classes \cite{Texas}. Scalar--Tensor theories \cite{10sotiriou}, higher dimensions theories (Brane--World 
scenarios\cite{14sotiriou} and Kaluza--Klein theory\cite{Kaluza}) and theories with a modified Lagrangian ($f(\az, \az_{ab}, \cdots)$) in both Palatini and metric--affine formulations \cite{sotiriou, carroll}, 
are some of them. 

In the Palatini gravity, the metric and the connection are considered as different degrees of freedom, but the matter Lagrangian does not depends on the connection. For the case of the standard  gravitational Lagrangian, $\sqrt{-g}\az$, this theory is identical with the Einstein's theory. If the matter Lagrangian depends on the connection then the theory is called metric--affine gravity.
Two special cases in which the Lagrangian is  $\sqrt{-g}f(\az)$ (non-linear Ricci scalar) and $\sqrt{-g}\az+\sqrt{-g}f(\az^{ab}\az_{ab})$ (non-linear Ricci squared) are considered frequently in the literature and have some interesting results as well as some drawbacks \cite{sotiriou, 3p}. 

Since in such theories the connection is different from the Christoffel symbols, one can introduce two different local frames. One in which the connection is locally zero and the other in which the Christoffel symbols are locally vanishing. 
These two frames are conformally related for Palatini $f(\az)$ gravity and thus the causal structure of both frames is the same, but this does not mean that they are completely equivalent from the physical point of view. We shall discuss about this point later. 
For Palatini $f(\az^{ab}\az_{ab})$ gravity, these are not conformally related and thus the causal structure of these two frames is different.
Investigation of the speed of light in these two frames helps one to choose one of these local frames as the local inertial frame.

In this paper after a brief review of the Palatini  $f(\az)$ and $f(\az^{ab}\az_{ab})$ gravity, we shall discuss about two possible local frames and about choosing one as the local inertial frame. Then,  we shall evaluate the different velocities of light, including the gravitational and electromagnetic wave velocities, and the space--time causal structure constant. 

It has to be noted here that according to Ostrogradski theorem\cite{Ostro}, higher derivative Lagrangians may suffer from instability problems. According to Ostrogradski, Hamiltonians associated with Lagrangians containing more than first time derivative would have linear instabilities, provided that higher derivatives cannot be eliminated by partial integration. It can be shown such instabilities does not exist for $f(\az)$ metric theories\cite{wood}.

In the Palatini formulation of $f(\az)$ and $f(\az^{ab}\az_{ab})$ theories, the Lagrangian depends only upon the first derivative of the fields and thus the result of Ostrogradski theorem is not applicable here.

This is the first reason why the metric formulation is not used here. There is also another reason. In the metric formulation, since the connection is equal to the Christoffel symbols there is only one local frame. This, as it would be clear at the end of this paper, leads to a constant speed of light. 
\section{Non-linear Ricci scalar and Ricci squared Palatini gravity in a nutshell}
The astronomical evidence for an accelerating cosmic expansion has stimulated many investigations into the other possible deviant gravitational effects, which might be responsible for this unexpected dynamics\cite{5}. Many investigators have also focused their attentions on modifying general relativity in the large scales. One example of them is provided by $f(\az)$ gravity models and the other example is    $f(\az, \az^{ab})$ gravity models.

When  the Einstein-Hilbert action is modified by adding some general function of $\az$ or $\az^{ab}\az_{ab}$, it becomes necessary to distinguish between two different variational approaches for deriving the field equations. In the metric approach, as in Refs \cite{6,7}, the metric components $g_{ab}$ are the only dynamical quantities and the field equations are generally of fourth-order. 
Within the Palatini variational approach, on the other hand, one assumes that the space--time is described 
by two independent geometrical objects, the metric and the connection. The first one defines  the causal structure of the space--time and the latter defines the affine structure of the manifold.  As a result the equations of motion are of the second order. 
\subsection{Palatini non-linear Ricci scalar gravity}
Let us first start with the non-linear Ricci scalar gravity in the Palatini formalism. The Lagrangian density of the theory is chosen to be an arbitrary function of the scalar curvature, $f(\az)$. According to the above discussions, the action of the Palatini $f(\az)$ theory is:
\begin{equation}
{\mathcal A}=\frac{1}{2\kappa}\int d^4x\sqrt{-g}f(\az[g,\Gamma])+{\cal A}_m
\label{action}
\end{equation}
where $\kappa=8\pi G/c_0^4$, and ${\mathcal A}_m$ is the matter action. Note that in the Palatini formalism the matter action does not depend on the connection. 
Since the affine connection is different from the Christoffel symbols, we can define two kinds of derivatives as:
\begin{equation}
\nabla_{a}X^{b} = \partial_{a}X^{b} + {b\brace ca}X^{c}
\end{equation}
\begin{equation}
D_{a} X^{b}= \partial_{a}X^{b} + \Gamma^{b}_{ca}X^{c}
\end{equation}

Variation of the action (\ref{action}) with respect to the metric ($g_{ab}$) and the connection ($\Gamma^{a}_{bc}$) leads to:
\[
\delta{\mathcal A}=\frac{1}{2\kappa}\int d^4x\sqrt{-g}\left [\left ( f' \az_{ab}-\frac{1}{2} f g_{ab} -\kappa T_{ab}\right ) \delta g^{ab}+\right .
\]
\begin{equation}
\left .
f' g^{ab}\left ( D_{c} \delta \Gamma^{c}_{ab}-D_{b}\delta\Gamma^{c}_{ac}\right ) \right ] 
\label{varac}
\end{equation}
in which a prime denotes differentiation with respect to $\az$ and $T_{ab}$ is the matter energy-momentum tensor defined as:
\begin{equation}
\delta{\mathcal A}_m=-\frac{1}{2}\int d^4x \sqrt{-g} T_{ab}\delta g^{ab}
\label{tt}
\end{equation}
Considering the metric and the connection as independent variables one arrives at the following equations of motion:
\begin{equation}
f'(\az)\az_{ab}-\frac{1}{2}f(\az)g_{ab}=\kappa T_{ab}
\label{eq1}
\end{equation}
and
\begin{equation}\label{eq01}
D_{c}\left (\sqrt{-g}f'g^{ab}\right )=0
\end{equation}

The last equation shows that making a conformal transformation as:
\begin{equation}
h_{ab}=f'g_{ab}
\end{equation}
leads to the fact that the connection is compatible with the metric $h_{ab}$ given by:
\begin{equation}\label{ggama}
\Gamma^{c}_{ab}= {c\brace ab}+ \gamma^{c}_{ab}={c\brace ab}+\frac{1}{2f'}\left [
2\delta^{c}_{(a}\partial_{b)}f'-g_{ab}g^{cd}
\partial_{d} f'\right ]
\end{equation}

A very interesting result can be obtained by investigation of the general covariance of the action   (\ref{action}). Let us make an infinitesimal transformation of the coordinates $x^{a}\rightarrow x^{a}+\epsilon^{a}$. Using the relation (\ref{tt}) and the fact that under coordinate transformation we have  $\delta g^{ab}=g^{c(a}\partial_{c}\epsilon^{b)}$, after some ordinary calculations, one finds that the corresponding N\"oether equation is:
\begin{equation}
\nabla_{a} T^{ab}=0
\label{t}
\end{equation}
For an explicit derivation of this relation the reader is referred to \cite{tomi}. The above equation means that the covariant divergence of the energy--momentum tensor does not vanish. This is not a bad signal since one can argue that it should be so. Any conservation relation should be of the form $\nabla_{a} j^{a}\equiv\frac{1}{\sqrt{-g}}\partial_{a}(\sqrt{-g}j^{a})=0$, to lead to the conservation of $\int d^3x\sqrt{-g}j^{a} n_{a}$ (with $n_{a}$ a time-like vector). Note that the correct measure for this integral is $\sqrt{-g}$ and not $\sqrt{-h}$, since this is $g_{ab}$ which is used to measure distance, area and volume, as we are dealing with a space--time structure in which the metric is $g_{ab}$ and the connection is $\Gamma^{c}_{ab}$. Extension to the case of higher rank tensors is straightforward.
\subsection{Palatini non-linear Ricci squared gravity}
In order to make a more general extension to the standard gravity one may add to the Einstein--Hilbert action a function of both the Ricci scalar and the Ricci tensor. A simple case that we shall use here is that this extra term is only a function of Ricci tensor, i.e.:
\begin{equation}
\mathcal{A} = \int d^{4}x \sqrt{-g}\left[{\az+f(\az^{ab}\az_{ab})\over 2\kappa} + {\mathcal{L}}_{m}\right]
\label{1}
\end{equation}
By varying the action (\ref{1})
we have
$$\delta \mathcal{A} = \int d^{4}x \frac{1}{2\kappa}\sqrt{-g} \left(\az_{ab} + 2F\az_{a}^{c}\az_{bc} - \frac{1}{2}g_{ab}
[\az + f(\az^{ab}\az_{ab})] - \kappa T_{ab}\right) \delta g^{ab} +$$
\begin{equation}
 \int d^{4}x \frac{1}{2\kappa}\sqrt{-g} (g^{ab} + 2F\az^{ab})[D_{c} \delta\Gamma^{c}_{ab} - D_{b} \delta\Gamma^{c}_{ac}]
\end{equation}
where $F = \partial f/\partial S$ with $S = \az^{ab}\az_{ab}$.
Therefore the \textit{modified Einstein equations} is:
\begin{equation}
\az_{ab} + 2F\az_{a}^{c}\az_{bc} - \frac{1}{2}g_{ab} [\az + f] = \kappa T_{ab}
\label{4}
\end{equation}
 
On the other hand, variation of the action with respect to the $\Gamma^{a}_{bc}$ leads to:
\begin{equation}
D_{e}\left[\sqrt{-g} (g^{ab} + 2Fg^{ac}\az_{cd}g^{bd})\right] = 0
\label{5}
\end{equation} 
Just like in the Palatini $f(\az)$ models, this equation implies some relation between the physical metric $g_{ab}$ and a new metric ($\sz_{ab}$) whose Christoffel symbol is $\Gamma^{a}_{bc}$. 

Since finding the solutions of the eqs.(\ref{4}), (\ref{5}) and  then finding the relation between $\sz_{ab}$ and $g_{ab}$ is generally a difficult task, one can use the covariant $1 + 3$ approach (see \cite{8,9,10}) at least for the cosmological solutions. 

Let's to assume that there is a family of preferred worldlines representing the average motion of matter at each point (as is the case for cosmology). This is the $4-$velocity $u^{a}= \frac{dx^{a}}{d\tau}$ normalized as $u^{a}u_{a} = 1$.
We assume this $4-$velocity is unique: that is, there is a well-defined preferred motion of matter at each space-time event. Given this, two  projection tensors can be defined. The first is $U^{a}_{b}=u^{a} u_{b}$ which projects anything parallel to the $4-$velocity vector, and the second is $h_{ab} =g_{ab}- u_{a} u_{b}$, which determines the (orthogonal) metric properties of the instantaneous rest-spaces of observers moving with $4-$velocity $u^{a}$. 

The energy-momentum tensor $T_{ab}$ can be decomposed relative to $u_{a}$ in the form
\begin{equation}
T_{ab} = \rho u_{a}u_{b} +2q_{(a}u_{b)} - ph_{ab} + \pi_{ab}
\label{6}
\end{equation}
in which , $\pi_{ab} = h_{a}^{c}h_{b}^{d}T_{cd} + ph_{ab}$ is the projected symmetric trace free anisotropic pressure(stress), $\rho = T_{ab}u^{a}u^{b}$ is the relativistic energy density relative to $u^{a}$, $q_{a} = h_{a}^{d}u^{c}T_{cd}$ is the relativistic momentum density and $p = \frac{-1}{3}h^{ab}T_{ab}$ is the isotropic pressure.

In a similar way  $\az_{ab}$ can be written as:
\begin{equation}
\az_{ab} = \Delta u_{a}u_{b} + \Xi h_{ab} + 2u_{(a}\Upsilon_{b)} + \Sigma_{ab}
\label{7}
\end{equation}  
Substituting eqs.(\ref{6}), (\ref{7}) into eq.(\ref{4}), we get the following four equations, that determines the coefficients
\begin{equation}
\Delta + 2F\Delta^{2} - \frac{1}{2}(\Delta + 3\Xi + f) = \kappa\rho
\label{8}
\end{equation}
\begin{equation}
\Xi +2F\Xi^{2} - \frac{1}{2}(\Delta + 3\Xi + f) = -\kappa p
\label{9}
\end{equation}
\begin{equation}
[1 + 2F(\Delta + \Xi)] \Upsilon_{a} = \kappa q_{a}
\label{eq10}
\end{equation}
\begin{equation}
(1 + 4F\Xi) \Sigma_{ab} = \kappa\pi_{ab}
\label{eq11}
\end{equation}
where f, F are functions of $\az^{ab}\az_{ab} = \Delta^{2} + 3\Xi^{2}$.Thus given the form of $f$ and the values of $\rho, p, q_{a}, \pi_{ab}$ the quantities $\Delta, \Xi, \Upsilon_{a}, \pi_{ab}$, can be obtained from the above equations at least numerically. So we have $\az_{ab}$ from eq.(\ref{7}). 

Now we should solve the eq.(\ref{5}). As it is stated previously one can introduce a new metric $\sz_{ab}$ by means of 
the following relation
\begin{equation}
\sqrt{-\sz} \sz^{ab} = \sqrt{-g} (g^{ab} + 2Fg^{ac} \az_{cd} g^{bd})
\end{equation}
After some calculations we can find the relations between the two metrics $\sz_{ab}$ and $g_{ab}$ and their inverses as
\begin{equation}
\sz_{ab} = \lambda g_{ab} + \xi_{ab},
\label{12}
\end{equation}
\begin{equation}
\sz^{ab} = \frac{1}{\lambda}g^{ab} + \zeta^{ab}
\label{13}
\end{equation}
where
\begin{equation}
\lambda = \sqrt{(1+2F\Delta)(1+2F\Xi)}
\label{14}
\end{equation}
\begin{equation}
\omega = \frac{1+2F\Xi}{1+2F\Delta}
\label{15}
\end{equation}
\begin{equation}
\xi_{ab} = \lambda(\omega - 1)u_{a}u_{b} - 4\sqrt{\omega}Fu_{(a}\Upsilon_{b)} - \frac{2F}{\sqrt{\omega}}\Sigma_{ab}
\label{16}
\end{equation}
\begin{equation}
\zeta^{ab} = \frac{1}{\lambda}\left(\frac{1}{\omega} - 1\right)u^{a}u^{b} + \frac{1}{\lambda^{2}}\frac{2F}{\sqrt{\omega}}[2u^{(a}\Upsilon^{b)} + \Sigma^{ab}]
\label{17}
\end{equation}
The relation between two different Christoffel symbol is 
\begin{equation}
\Gamma^{a}_{bc}={a\brace bc}+ \gamma^{a}_{bc}
\label{18}
\end{equation}
where
$$\gamma^{a}_{bc} \equiv \frac{1}{2\lambda}[\delta^{a}_{b}\nabla_{c}\lambda + \delta^{a}_{c}\nabla_{b}\lambda - g_{bc}\nabla^{a}\lambda] + \frac{1}{2}[\zeta^{a} _{b}\nabla_{c}\lambda + \zeta^{a}_{c}\nabla_{b}\lambda - g_{bc}\zeta^{ad}\nabla_{d}\lambda]$$
\begin{equation}
\ \ \ \ + \frac{1}{2\lambda}[\nabla_{b}\xi^{a}_{c} + \nabla_{c}\xi^{a}_{b} - \nabla^{a}\xi_{bc}] + \frac{1}{2}\zeta^{ad}[\nabla_{c}\xi_{bd} + \nabla_{b}\xi_{cd} - \nabla_{d}\xi_{bc}] 
\label{19}
\end{equation}
and also we have
\begin{equation}
\az_{ab} = R_{ab} + \nabla_{c}\gamma^{c}_{ab} - \nabla_{b}\gamma^{c}_{ac} + \gamma^{d}_{ab}\gamma^{c}_{cd} - \gamma^{d}_{ac}\gamma^{c}_{bd} 
\label{20}
\end{equation}
in which $R_{ab}$ is the Ricci tensor with respect to Christoffel symbols. 
For more details see \cite{3,3p}.

For our later use, we can write the modified Einstein eq.(\ref{4}) as
\begin{equation}
R_{ab} - \frac{1}{2}g_{ab}R = \kappa T_{ab} + \kappa T_{ab}^{eff}
\label{21}
\end{equation}  
where
\begin{equation}
\kappa T_{ab}^{eff} \equiv \frac{a}{2}g_{ab}(f + g^{cd} \Pi_{cd}) - \Pi_{ab} - 2F\az_{a}^{c}\az_{cb}
\label{22}
\end{equation}
and
\begin{equation}
\Pi_{ab} = \nabla_{c}\gamma^{c}_{ab} - \nabla_{b}\gamma^{c}_{ac} + \gamma^{d}_{ab}\gamma^{c}_{cd} - \gamma^{d}_{ac}\gamma^{c}_{bd} 
\label{23}
\end{equation}

As a special case for the FRW cosmological models we have $q^{a} = \pi^{ab} = 0$ so that $T_{ab} = \rho u_{a}u_{b} - ph_{ab}$. According to this and eqs.(\ref{eq10}), (\ref{eq11}) we have $\Upsilon_{a} = \Sigma_{ab} = 0$. The remaining parameters satisfy the following equations:
\begin{equation}\label{delta}
\Delta + 2F\Delta^{2} - \frac{1}{2}(\Delta + 3\Xi + f) = \kappa\rho
\end{equation}
\begin{equation}\label{xi}
\Xi + 2F\Xi^{2} - \frac{1}{2}(\Delta + 3\Xi + f) = -\kappa p
\end{equation}
and 
\begin{equation}
\xi_{ab} = \lambda(\omega - 1)u_{a}u_{b},\ \ \ \ \ \zeta^{ab} = \frac{1}{\lambda}(\frac{1}{\omega}-1)u^{a}u^{b}
\end{equation}
So we have
\begin{equation}
\sz_{ab} = \lambda g_{ab} + \lambda(\omega - 1)u_{a}u_{b}
\label{24}
\end{equation}
\begin{equation} 
\sz^{ab} = \frac{1}{\lambda}g^{ab} + \frac{1}{\lambda}(\frac{1}{\omega} - 1)u^{a}u^{b}
\end{equation}

It should be noted at this end that the conservation relation of the energy--momentum  tensor is again of the form $\nabla_a T^{ab}=0$. This is because of the fact that the gravitational and matter parts of the action functional are scalars under general coordinate transformations. And as it is proved previously this leads to the above conservation law.
\section{Local inertial frame}
In order to discuss some kinds of the  speed of light, we need to go to the local frame. For example for defining $c_{ST}$ one should examine the local causal structure. Also to separate the gravitational effects from the effects caused by variation of the speed of light in the $c_{EM}$ it is helpful to go to the local inertial frame.

In the Einstein's theory of  gravity one has the equivalence principle which asserts that there is some local frame in which gravity is locally absent. This is hard-coded in the theory such that the local frame in which the metric is Minkowskian has zero connection (the Christoffel symbols). In the Palatini formalism as while as $f(\az)=\az$, the same is true, but for a general $f(\az)$ this fails. That is to say, one is able to define two local frames. In the first one,  metric is locally Minkowskian, but the connection is not zero. In fact we have: $g_{ab}=\eta_{ab}$, ${c\brace ab}=0$ and $\Gamma^{c}_{ab}=\gamma^{c}_{ab}$. There is also a second local frame in which the connection is zero and thus: $g_{ab}=\frac{1}{f'}\eta_{ab}$, ${c\brace ab}=-\gamma^{c}_{ab}$ and $\Gamma^{c}_{ab}=0$.
In order to understand the physical significance of the above mentioned frames, let's to investigate the trajectory of a test particle in these two frames. To do so, we consider a dust with $T_{ab}=\rho u_{a} u_{b}$, where $\rho$ is particle density and $u_{a}$ is the velocity four--vector. The conservation relation (\ref{t}) leads to:
\begin{equation}
\nabla_{a}(\rho u^{a} u^{b})=0
\end{equation}
and we have to consider the particle conservation law, too:
\begin{equation}
\nabla_{a}(\rho u^{a})=0
\end{equation}
Note that according to the discussion of the previous section, this conservation equation should be evaluated by $\nabla$. Combination of these two equations leads to the trajectory of a test particle which is $u^{a}\nabla_{a} u^{b}=0$, or:
\begin{equation}\label{geodesic eq}
\frac{d^2x^{a}}{ds^2}+{a\brace bc}\frac{dx^{b}}{ds}\frac{dx^{c}}{ds}=0
\end{equation}
In the first local frame, the particle trajectory is a straight line 
\begin{equation}
\frac{d^2x^{a}}{ds^2}=0
\end{equation}
while in the second frame one has
\begin{equation}
\frac{d^2x^{a}}{ds^2}-\frac{dx^{a}}{ds}\frac{d\ln f'}{ds}+\frac{1}{2}\eta^{ab}\partial_{b} f'=0
\label{eq88}
\end{equation}

Note that the same is true for non-linear Ricci squared theories, except that the form of $\gamma^a_{bc}$ is different. Therefore, generally the trajectory of a test particle in the second frame is
\begin{equation}
\frac{d^{2}x^{a}}{ds^{2}}-\gamma^{a}_{bc}\frac{dx^{b}}{ds}\frac{dx^{c}}{ds}=0
\label{ao1}
\end{equation} 

Now the question is that which one of these frames is the local inertial frame? 
It has to be noted that although in the first frame the particle moves on a straight line but this does not mean that equivalence principle is satisfied since gravity is present via the non--vanishing connection in other physical relations. Although we are talking about Palatini formalism in which the matter action is not coupled to the connection, but this is not the whole story.

In Palatini $f(\az^{ab}\az_{ab})$ and $f(\az)$ gravity we assumed that the space--time structure is defined via the metric (as the device for defining length) and the connection (as the device for parallel transporting). This means that in any experiments that the concept of parallelism is important the effect of the connection can be seen. 

For example suppose that we want to measure the focal length of a lens in the first local frame. To do so we need to send at least two rays of light towards the lens. Although in this frame the Maxwell equations has no dependence on the connection and thus the light rays behave like in special relativity, but to make them parallel we need to use the connection which is not zero. In other words in the first local frame gravity is present in defining the focal length of the lens. The distance between geodesics and the description of a congruence of them are of course given by the affine connection. 

To state this important point in another way, let us to stress that the
theory is not only the field equations. It is the field equations derived from the action defined on some spacetime with predefined properties. For the Palatini theory one assumes the spacetime has an independent affine connection and thus any parallel transport should be evaluated using it.
 So one expects to have changes in the geometrical concepts like geodesic deviation and Raychaudhuri's equation representing how a flux of geodesics expands. In this theory the existence of two different connection fields has new consequences. The geodesics are determined by the Christoffel symbols (this choice is motivated by energy-momentum conservation) but the equation that governs the evolution of the deviation vector involves the affine connection (motivated by the fact that the covariant derivative or parallel propagation along any arbitrary curve is defined by the affine connection). For more details see \cite{shojai}.
 
Therefore it seems that the first local frame is not the local inertial frame.
On the other hand, in the second frame the space--time for Ricci scalar model  is locally conformally flat and the connection is zero and for Ricci squared model only the connection is zero. Since the Christoffel symbols are not zero, matter fields and the test particle trajectory are not the special relativistic ones, it is given by equation (\ref{ao1}). In order to understand physical implications of this equation, it is instructive to define another concept of the speed of light, the speed used to synchronize the clocks. In the standard relativity, the synchronization procedure is as follows. In the local inertial frame, point 1 sends an information signal (with velocity $c_C=c_0$) to point 2 at a distance $d\ell$ from point 1. Then if the observer at point 2 adjust its clock $d\ell/c_0$ ahead, the two points are synchronized.
As a result the proper time in this frame is defined as $ds=c_Cd\tau$. It is now instructive to write the test particle trajectory (equation (\ref{ao1})) in terms of the proper time ($d\tau=ds/c_C$):
\begin{equation}
\frac{d^2x^a}{d\tau^2}-\gamma^a_{bc}\frac{dx^b}{d\tau}\frac{dx^c}{d\tau}-\frac{dx^a}{d\tau}\frac{1}{c_C}\frac{dc_C}{d\tau}=0
\end{equation}

Therefore, although in this frame the connection is zero and thus can be interpreted as the local inertial frame, but the particle's trajectory is not a line. One can identify the extra terms with the force due to variation of the speed of light. To see this, consider the case of Palatini $f(R)$ gravity, in the second local frame. The observer at point 2 has to adjust it clock ahead with the amount $\sqrt{f'}d\ell/c_0$ (with $d\ell^2=f'|d\vec{x}|^2$). This show that for synchronization one should use the velocity $c_C=c_0/\sqrt{f'}$.
Therefore the test particle motion is given by:
\begin{equation}
\frac{d^2x^a}{d\tau^2}+\frac{dx^a}{d\tau}\frac{1}{c_C}\frac{dc_C}{d\tau}-\eta^{ab}c_0^2\frac{\partial_bc_C}{c_C}=0
\end{equation}
It is clear now that the extra terms are forces exerted on the particle because of the variation of the clock synchronization speed of light. A similar discussion can be done for non-linear Ricci squared theories. 
\section{Speed of Light}
Let us now consider the meaning and the properties of the speed of light in the framework of the geometrical structure of the Palatini $f(\az)$ and $f(\az^{ab}\az_{ab})$ gravity theories. As it is discussed earlier, one should distinguish between different velocities of light.  
\subsection{The space--time--matter coupling constant}
As it is clear from the actions chosen for both nonlinear Ricci scalar and Ricci squared Palatini theories, we have assumed that $c_E=c_0$. Any attempt to make this velocity varying should be done by making it dynamical and adding dynamical terms for it to the action. This leads to some scalar-tensor theory and is discussed in the literature\cite{5khodam,ellis,oooo}.
\subsection{The gravitational wave velocity}
In order to drive the gravitational wave velocity, one needs to linearize the field equations around the no--gravity solution. Since it is not clear what is and whether the linearized solution exists for a general $f(\az)$ Palatini theory\cite{sotiriou, sotilin, olmo}, we restrict ourself to the case in which $f(\az)$ has not negative powers of the Ricci scalar. That is we assume $f(\az)=\az+a_1\az^2+a_2\az^3+\cdots$ where $a_i$'s are constants.
For $f(\az)$ theory, taking the trace of equation(\ref{eq1}), we get
\begin{equation}
f'(\az)\az-2f(\az) = \kappa T
\end{equation}
Assuming $T=0$, which includes   
vacuum and electro-vacuum, for a given $f$, this is an algebraic equation for $\az$. The Ricci scalar will therefore be a constant and a root of the equation
\begin{equation}\label{sot}
f'(\az_0)\az_0-2f(\az_0) = 0
\end{equation}
We will not consider cases for which this equation has no roots since it can be shown that the field equations are then not consistent.
As a result according to the field equations, 
the theory reduces to \textit{General Relativity } with a \textit{cosmological constant} given by:
\begin{equation}
\Lambda= \frac{\az_{0}}{4}
\end{equation}
For more details see \cite{sotiriou}.
Therefore investigation of gravitational waves in this case is not different from the standard gravity and thus the gravitational wave velocity is just $c_{0}$.

Now we are going to calculate the gravitational wave velocity for Ricci squared gravity. For Ricci squared theory, again we assume that we have not negative powers of Ricci squared. That is:
\begin{equation}
f(\az_{ab}\az^{ab})=a_1 \az_{ab}\az^{ab} + a_2 (\az_{ab}\az^{ab})^2+\cdots
\end{equation}
In order to linearize the equations (\ref{4}) and (\ref{5}) we begin by assuming that the metric differs only slightly from the Minkowskian metric, that is
\begin{equation}\label{g}
g_{ab} = \eta_{ab} + \varepsilon \tilde{g}_{ab}+{\cal O}(\varepsilon^{2})
\end{equation}
where $\varepsilon$ is a small dimensionless parameter and, throughout, we shall neglect terms of second order or higher in $\varepsilon$.
One can easily verify that in the zeroth order the affine connection is zero, and thus
\begin{equation}
\Gamma^{a}_{bc} = 0 + \varepsilon \tilde{\Gamma}^{a}_{bc}+{\cal O}(\varepsilon^{2})
\end{equation} 
Neglecting terms of ${\cal O}(\varepsilon^{2})$, the Ricci tensor is
\begin{equation}
\az_{ab} = \varepsilon ( \tilde{\Gamma}^{c}_{ab,c} - \tilde{\Gamma}^{c}_{ac,b})
\end{equation}
and the Ricci scalar is
\begin{equation}\label{R}
\az = g^{cd} \az_{cd} = \varepsilon\eta^{cd} (\tilde{\Gamma}^{f}_{cd,f} - \tilde{\Gamma}^{f}_{cf,d})
\end{equation}
Finally the linearized modified Einstein equation(\ref{4}) gives (up to ${\cal O}(\varepsilon^{2})$ corrections):
\begin{equation}
\tilde{\Gamma}^{c}_{ab,c} - \tilde{\Gamma}^{c}_{ac,b} - \frac{1}{2} \eta_{ab}\eta^{cd}(\tilde{\Gamma}^{f}_{cd,f} - \tilde{\Gamma}^{f}_{cf,d}) = 0
\label{gauge}
\end{equation}

We may seek for wave solutions of the above equation by assuming the relations
\begin{equation}\label{connection}
\tilde{\Gamma}^{a}_{bc} = \xi^{a}_{bc} \ \  e^{ik\cdot x}
\end{equation}
\begin{equation}\label{metric}
h_{ab} = \zeta_{ab} \ \  e^{ik\cdot x}
\end{equation}
where $\xi^{a}_{bc}$ is symmetric in lower indices because of the fact that we have assumed space--times without torsion. Putting equation (\ref{connection}) into equation (\ref{gauge}), we get the following  \textit{gauge condition}: 
\begin{equation}
k_{c}\xi^{c}_{ab} - k_{b}\xi^{c}_{ac} = 0
\end{equation} 

The wave equation is derived via linearization of equation (\ref{5}). It gives:
\begin{equation}
\partial _{e} (\sqrt{-g}g^{ab}) + \varepsilon \tilde{\Gamma}^{a}_{ec}\eta^{cb} + \varepsilon \tilde{\Gamma}^{b}_{ec}\eta^{ac} + 2a_1\varepsilon\eta^{ac}\eta^{bd}\partial_{e} (\tilde{\Gamma}^{f}_{cd,f} - \tilde{\Gamma}^{f}_{cf,d}) = 0
\label{wave}
\end{equation}
Using the equation $\nabla_{e}(\sqrt{-g}g^{ab}) = 0$, we can simplify the first term and thus we have:
$$ -\tilde{g}^{ab}_{,e} - \frac{1}{2}\tilde{g}^{a,b}_{e} - \frac{1}{2}\tilde{g}^{b,a}_{e} + \frac{1}{2}\tilde{g}_{e}^{b,a} + \frac{1}{2}\tilde{g}_{e}^{a,b}$$ 
\begin{equation}
\eta^{cb}\tilde{\Gamma}^{a}_{ec} + \eta^{ac}\tilde{\Gamma}^{b}_{ec} + 2a_1\eta^{ac}\eta^{bd}\partial_{e}(\tilde{\Gamma}^{f}_{cd,f}  -  \tilde{\Gamma}^{f}_{cf,d}) = 0
\end{equation} 
By using the equations (\ref{connection}) and (\ref{metric}) and  the above gauge condition, one gets:
\begin{equation}
ik_{c}\zeta^{ab}= \eta^{ad}\xi^{b}_{cd} + \eta^{bd}\xi^{a}_{cd}
\end{equation}
After multiplying both sides of the above equation by $-ik^c$, we find out that the \textit{wave equation} becomes
\begin{equation}
k^{2}\zeta^{ab}= -ik^c\left ( \eta^{ad}\xi^{b}_{cd} + \eta^{bd}\xi^{a}_{cd}\right )
\label{wave equation}
\end{equation}
Taking into account the fact that Palatini theory is invariant under general covariance and the so-called projection transformation defined as $\Gamma^a_{bc}\rightarrow\Gamma^a_{bc}+\delta^a_b\Lambda_c+
\delta^a_c\Lambda_b$ with $\Lambda_b$ an arbitrary four vector, we have an 16-fold freedom. Therefore choosing $k^c\xi^a_{bc}=0$, we get $k^2=0$. As a result the gravitational wave velocity in the nonlinear Ricci squared theory is equal to $c_0$.
\subsection{The electromagnetic wave velocity}
Let us now consider the case of electromagnetic waves by choosing:
\begin{equation}
{\cal A}_m=-\frac{1}{4}\int d^4x \sqrt{-g} F^{ab}F_{ab}
\end{equation}
where the electromagnetic field is defined as
\begin{equation}
F_{ab}=D_{a}A_{b} - D_{b}A_{a} = \nabla_{a}A_{b}-\nabla_{b}A_{a}= \partial_{a}A_{b} - \partial_{b}A_{a}
\label{eq99}
\end{equation}
The field equations are thus:
\begin{equation}
\nabla_{a} F^{ab}=0
\end{equation}
To find the electromagnetic wave velocity, we consider a solution of type $A_{a}=\epsilon_{a}\exp (ik\cdot x)$. 
In order to distinguish between the gravitational effects from the varying speed of light let's to consider the wave velocity in the two possible local frames.

In the first local frame in which the metric is Minkowskian, for both Ricci scalar and Ricci squared gravity we get:
\begin{equation}
\epsilon \cdot k=0
\end{equation}
and
\begin{equation}
k^2=0
\end{equation}
The first equation represents gauge invariance and it is equivalent to $\nabla_{a}A^{a} = 0$ and the second shows that $c_{EM}=c_0$ in this frame. But remember that in this frame the gravitational effects are present.

In the second local frame for which the connection is zero, there is a distinction between Ricci scalar and Ricci squared gravity. For Ricci scalar model one has:
\begin{equation}
\epsilon^{a}(k_{a}+2i\partial_{a}\ln f')=0
\end{equation}
and
\begin{equation}
k^2+2ik^{a} \partial_{a}\ln f'=0
\label{sss}
\end{equation}
The first equation is gauge invariance ($\nabla_{a}A^{a} =0$), and the second shows that generally $k^2\neq 0$. Since the extra term is imaginary and is proportional to the wave vector one has both attenuation of the wave and that $c_{EM}\neq c_0$. In fact this extra term is not a constant and thus the electromagnetic wave velocity is varying.

For Ricci squared gravity one has
\begin{equation}
\epsilon^{a}(k_{a}+i\gamma^{c}_{ca})=0
\end{equation}
and
\begin{equation}
k^{2}+i\gamma^{a}_{ac}k^{c}=0
\label{dis}
\end{equation}
Again the first equation represents the gauge invariance and the second one introduces a varying speed for the electromagnetic waves.

In order to see the physical implications of this last equation, let us evaluate it for FRW cosmological model. A look at equation (\ref{19}) shows that for such a model $\gamma^a_{ac}=\frac{2}{\lambda}\partial_c\lambda+\frac{1}{2\omega}\partial_c\omega$. Thus the only non--zero element is $\gamma^a_{a0}\equiv q(t)$ which is some function of time. Assuming a plane wave moving in x direction with $\vec{k}=(\kappa+i\alpha)\hat{x}$ and using equation (\ref{dis}) one can easily calculate the phase velocity as:
\begin{equation}
\left ( \frac{c_{EM}}{c_0}\right )_{phase}=\frac{1}{\sqrt{1+\frac{q^2}{4\kappa^2}}}
\end{equation}
and the group velocity as:
\begin{equation}
\left ( \frac{c_{EM}}{c_0}\right )_{group}=\frac{1+\frac{q^2}{2\kappa^2}}{\left (1+\frac{q^2}{4\kappa^2}\right )^{3/2}}
\end{equation}
and also the attenuation factor as:
\begin{equation}
\alpha=\frac{q}{2}\frac{1}{\sqrt{1+\frac{q^2}{4\kappa^2}}}
\end{equation}
The same result is valid for Ricci scalar model as it is clear from equation (\ref{sss}). The change in the velocity of electromagnetic waves affects the way atoms radiate in the far galaxies, and therefore there is a contribution in the cosmological redshift from the varying speed of electromagnetic waves. 
\subsection{The space--time causal structure constant}
Since the two local frames are conformally related for the case of Palatini Ricci scalar theories, the local causal structure is identical to the Minkowskian case. This means that in both frames we $c_{ST}=c_0$. 

In the case of Palatini Ricci squared, it is clear that in the first local frame in which the metric is Minkowskian we have $c_{ST}=c_0$. On the other hand in the second local frame the space--time structure constant $c_{ST}$ has a varying character, because of the fact that metric is not locally Minkowskian. In order to see how it looks like, let us to investigate it for a cosmological model. We consider the FRW model. For such a simple model, we can follow the relations easily.
In the Robertson-Walker space-time, we have 
\begin{equation}
ds^{2} = g_{ab}dx^{a}dx^{b} = c_{0}^{2}dt^{2} - a^{2}(t) [\frac{dr^{2}}{1-kr^{2}} + r^{2}(d\theta^{2} + \sin^{2}\theta d\phi^{2})]
\label{25}
\end{equation}
and $u^{a}= (1,0,0,0)$. By using of the equations (\ref{24}) and  (\ref{25}) we can find $\sz_{ab}$
$$\sz_{00} = \lambda g_{00} + \lambda (\omega - 1) = \lambda\omega $$
$$\sz_{ij} = \lambda g_{ij}  \longrightarrow  \sz_{ii} = \lambda g_{ii}$$
$$\sz_{0i} = \lambda g_{0i} = 0$$
So we have
\begin{equation}
\sz_{ab} = diag \left( \lambda\omega , \frac{-\lambda a^{2}(t)}{1-kr^{2}} , -\lambda r^{2}a^{2}(t) , -\lambda r^{2}\sin^{2}\theta a^{2}(t) \right)
\label{26}
\end{equation}
In order to go to the first local frame for which  $g_{ab}$ is locally $\eta_{ab}$, one can use the relation
\begin{equation}
g^{ab} = e^{a}_{i}(x) e^{b}_{j}(x) \eta^{ij}
\end{equation} 
with the local tetrad fields given by:
\begin{equation}
e^{a}_{i}(x) = diag \left( 1, \frac{\sqrt{1-kr^{2}}}{a(t)} , \frac{1}{ra(t)} , \frac{1}{r\sin\theta a(t)} \right)
\label{27}
\end{equation}
It is quiet obvious that in this frame: 
\begin{equation}
c_{ST} = c_{0}
\label{28}
\end{equation} 

On the other hand, in the second local frame which is local inertial frame, we shall have:
\begin{equation}
\sz^{ab} = \tilde{e}^{a}_{i}\tilde{e}^{b}_{j}\eta^{ij}
\end{equation}
Investigation of equation (\ref{26}) shows that the tetrad fields are given by:
\begin{equation}
\tilde{e}^{a}_{i} = diag \left( \frac{1}{\sqrt{\lambda\omega}} , \frac{\sqrt{1-kr^{2}}}{\sqrt{\lambda}a(t)} , \frac{1}{\sqrt{\lambda}ra(t)} , \frac{1}{\sqrt{\lambda}rsin\theta a(t)} \right)
\label{32}
\end{equation}
In order to derive $c_{ST}$ in this frame, we have to obtain the form of the space--time metric using the relation:
\begin{equation}
g_{ab} \tilde{e}^{a}_{i}\tilde{e}^{b}_{j} = \tilde{g}_{ij}
\end{equation}
Therefore:
\begin{equation}
\tilde{g}_{ij} = diag ( \frac{1}{\lambda\omega} , -\frac{1}{\lambda} , -\frac{1}{\lambda} , -\frac{1}{\lambda} )
\end{equation}
The line element in this frame is thus:
\begin{equation}
ds^{2} = \frac{1}{\lambda} [\frac{1}{\omega}c_{0}^{2}dt^{2} - dr^{2} - r^{2}d\theta^{2} - r^{2}sin^{2}\theta d\phi^{2}]
\end{equation}
Assuming a radial null ray, we can find the signal speed which is $c_{ST}$: 
\begin{equation}
c_{ST} = \frac{1}{\sqrt{\omega}}c_{0}
\label{33}
\end{equation}
For a simple model for which $f(\az_{ab}\az^{ab})=a_1 \az_{ab}\az^{ab}$ we obtain
\begin{equation}
c_{ST} = \sqrt{\frac{1+2a_1\Delta}{1+2a_1\Xi}} c_{0}
\label{signal}
\end{equation}
Let us consider this equation for three situations; namely for dust, radiation and cosmological constant.
Using the relation $ \az^{ab}\az_{ab}=\Delta^{2} +3\Xi^{2}$ and the equations (\ref{delta}), (\ref{xi}) we get:
\begin{equation}
\Delta + 3a_1\Delta^{2} - 3\Xi - 3a_1\Xi^{2} = 2\kappa\rho
\end{equation}
and
\begin{equation}
-\Delta -a_1\Delta^{2} -\Xi +a_1\Xi^{2} = -2\kappa p
\label{55}
\end{equation}
For dust one has $p=0$ and $\rho\sim a^{-3}$, so 
\begin{equation}
c_{ST} \simeq (1 + a_1\kappa a^{-3}) c_0
\end{equation} 
And for radiation we have $\rho\sim a^{-4}$ and $p\sim \frac{1}{3}a^{-4}$, so that
\begin{equation}
c_{ST} \simeq (1 + \frac{4}{3}a_1\kappa a^{-4})c_0
\end{equation}
Finally for cosmological constant $\omega$ will be a constant so that the velocity of signal is a constant less that $c_0$.

Since the region that particles can have interaction with eachother is defined using $c_{ST}$, a varying $c_{ST}$ can have observable results on problems like the horizon problem.
\section{Conclusions}
In order to conclude let us to stress again that five different velocities can be distinguished when one wants to let the velocity of light to vary. In the standard theory all these five velocities are equal to the constant $c_0=3\times 10^8\ meters/second$. 

The first one appears in the coupling constant of gravity and matter and here we chose it to be constant. Calling it $c_E$ we have $c_E=c_0$.
 
The second velocity is the gravitational wave velocity $c_{GW}$. We assumed that no negative powers of the Ricci scalar or Ricci squared is present and then we have shown that $c_{GW}$ in both scalar and Ricci squared theories is equal to $c_0$. 

Since in the theories we concern here the connection is not the metric connection, two different local frames are distinguishable. The first one is the one in which the metric is locally Minkowskian. We saw that the trajectory of a test particle in this frame is a straight line. But this does not simply mean that this frame is the local inertial (no--gravity) frame. Because gravity arises as coupling of the connection to other physical equations and also via the concept of parallelism and in this frame the connection is not zero. 
In the second local frame in which the connection is zero, the test particle trajectory is not a line. We saw that one can assign the extra terms in the particle's trajectory equation as the force due to variation of the clock synchronization speed, the speed of the signal used to make faraway clocks synchronized.

The fourth speed is the electromagnetic wave velocity. We saw that for both Palatini non-linear Ricci scalar and Ricci squared cases, the velocity of electromagnetic waves is  simply $c_0$ in the first local frame. But in this frame gravity is present and this velocity is the velocity of electromagnetic waves modulated with gravitational effects. On the other hand in the second local frame which is the local inertial frame, the velocity of electromagnetic waves is not $c_0$. It is varying.

The fifth velocity appears as the space--time causal structure constant or the information velocity. 
In the Palatini $f(\az)$ theory since the two local frames are conformally related, we have $c_{ST}=c_0$ in both frames. But for the case of Palatini Ricci squared, only in the first local frame we have $c_{ST}=c_{0}$,. In the second local frame $c_{ST}$ is varying. We have investigated it for FRW cosmological model and shown that in this local frame that the information velocity is $\frac{1}{\sqrt{\omega}}c_{0}$. 

To sum up table (\ref{tab1}) may help. 

\begin{center} 
\begin{table}[h]
\begin{tabular}{|c|c|c|}
\hline  &Palatini $f(\az)$ gravity & Palatini $f(\az^{ab}\az_{ab})$ gravity  \\ 
\hline $c_E$ & $c_0$ & $c_0$ \\ 
\hline $c_{GW}$ & $ c_{0}$ & $c_{0}$ \\ 
\hline $c_{EM}$ & varying & varying \\ 
\hline $c_{ST}$ & $c_0$ & varying ($\frac{1}{\sqrt{\omega}}c_{0}$) \\ 
\hline $c_C$ & varying ($\frac{c_{0}}{\sqrt{f'}}$) & varying  \\ 
\hline 
\end{tabular} 
\caption{Comparison of different speeds of light in the Palatini $f(\az)$ and $f(\az^{ab}\az_{ab})$ theories.}
\label{tab1}
\end{table}
\end{center}

At this end it is clear that if we use the metric formulation, $c_{EM}$ and $c_{ST}$ would be constant and equal to $c_0$. This is because of the fact that in the metric formulation the connection and the Christoffel symbols are equal and thus there is only one local frame. In this local frame the metric is Minkowskian and thus $c_{ST}=c_0$ and also Maxwell equations have their special relativistic form so we have $c_{EM}=c_0$.

\textbf{Acknowledgments}: The authors wish to thank Fatimah Shojai for her very fruitful hints and discussions. 
This work is partly
supported by a grant from university of Tehran and partly by a grant
from center of excellence of department of physics on the structure
of matter.

\end{document}